%% file: Laclapiererevised12.tex
\documentstyle[agupp,psfig]{article}
\lefthead{ Helmstetter et al.}
\righthead{Slider-block Model for Landslides } 

 \authoraddr{Agn\`es Helmstetter, Laboratoire de G{\'e}ophysique
Interne et Tectonophysique,
Observatoire de Grenoble, Universit{\'e} Joseph Fourier, BP 53X,
38041 Grenoble Cedex, France. (e-mail: helmstet@moho.ess.ucla.edu)}
\input{update.tex}
\begin{document}

\title{\Large Slider-Block Friction Model for Landslides: Application
to Vaiont and La Clapi\`ere Landslides}

\author{\large A. Helmstetter$^{1}$, D. Sornette$^{2-4}$,
J.-R. Grasso$^1$, J. V. Andersen$^{2,5}$, S. Gluzman$^4$
and V. Pisarenko$^6$}
\noindent
$^1$ LGIT, Observatoire de Grenoble, Universit\'e Joseph Fourier, France\\
$^2$ LPMC, CNRS UMR 6622 and
Universit\'{e} de Nice-Sophia Antipolis\\ Parc Valrose, 06108 Nice, France\\
$^3$ Department of Earth and Space Sciences\\
   University of California, Los Angeles, California 90095-1567\\
$^4$ Institute of Geophysics and Planetary Physics\\
University of California, Los Angeles, California 90095-1567\\
$^5$ U. F. R. de Sciences Economiques, Gestion, Math\'ematiques et
Informatique, \\ CNRS UMR7536 and Universit\'e Paris X-Nanterre \\
92001 Nanterre Cedex, France\\
$^6$ International Institute of Earthquake Prediction Theory and
Mathematical Geophysics\\ Russian Ac. Sci. Warshavskoye sh., 79, kor. 2,
Moscow 113556, Russia

\newcommand{\be}{\begin{equation}}
\newcommand{\ee}{\end{equation}}
\newcommand{\ba}{\begin{eqnarray}}
\newcommand{\ea}{\end{eqnarray}}
\newenvironment{technical}{\begin{quotation}\small}{\end{quotation}}

\vskip -1cm

\begin{abstract}

Accelerating displacements preceding some catastrophic landslides 
have been found empirically to follow a time-to-failure power law, 
corresponding to a finite-time singularity of the
velocity $v \sim 1/(t_c-t)$ [{\it Voight}, 1988]. Here, we provide a
physical basis for this phenomenological law based on
a slider-block model using
a state and velocity dependent friction law established in the
laboratory and used to model earthquake friction. This physical model
accounts for and generalizes Voight's observation:
depending on the ratio $B/A$ of
two parameters of the rate and state friction law and on the initial
frictional state of the sliding surfaces characterized by a
reduced parameter $x_i$, four possible regimes are found.
Two regimes can account for an acceleration of the displacement.
For  $B/A>1$ (velocity weakening) and $x_i<1$,
the slider block exhibits an unstable acceleration
leading to a finite-time singularity of the displacement
and of the velocity $v \sim 1/(t_c-t)$, thus rationalizing  Voight's 
empirical law.
An acceleration of the displacement can also be reproduced in the
  velocity  strengthening regime, for $B/A<1$ and $x_i>1$. In this case,
the acceleration of the displacement evolves toward a stable sliding with
a constant sliding velocity.
The two others cases ($B/A<1$ and $x_i<1$, and $B/A>1$ and $x_i>1$) give
a deceleration of the displacement.
We use the slider-block friction model to analyze quantitatively the 
displacement
and velocity data preceding two landslides, Vaiont and La Clapi\`ere.
The Vaiont landslide was the catastrophic culmination of an accelerated slope
velocity. La Clapi\`ere landslide was characterized by a peak of slope acceleration
that followed decades of ongoing accelerating displacements, succeeded by a restabilizing phase.
Our inversion of the slider-block model on these
data sets shows good fits and suggest to classify the Vaiont
(respectively La Clapi\`ere) landslide as belonging to the velocity weakening
unstable (respectively strengthening stable) sliding regime.

\end{abstract}
\begin{article}

\section{Introduction}

Landslides constitute a major geologic hazard of strong concern in most
parts of the world. The force of rocks,
soil, or other debris moving down a slope can devastate anything in its
path. In the United States for instance,
landslides occur in all 50 states and cause \$1-2 billion in
damages and more than 25 fatalities on average each year. The situation
is very similar in costs and casualty rates in the European Union.
Landslides occur in a wide variety of geomechanical contexts,
geological and structural settings, and as a response to
various loading and triggering processes. They are often associated with
other major natural disasters such as
earthquakes, floods and volcanic eruptions.

Landslides sometimes strike without discernible warning.
There are however well-documented cases of precursory signals, showing
accelerating slip over time scales of weeks to decades
(see [{\it Voight (ed)}, 1978] for a review).
While only a few such cases have been monitored in the past,
modern monitoring techniques are bound to provide a wealth of new
quantitative observations based on GPS and SAR (synthetic
aperture radar) technology to map the surface
velocity field [{\it Mantovani et al.}, 1996; {\it Fruneau et al.}, 1996;
{\it Parise}, 2001; {\it Malet et al.}, 2002] and seismic
monitoring of slide quake activity [{\it Gomberg et al.}, 1995;
{\it Xu et al.}, 1996; {\it Rousseau}, 1999;
{\it Caplan-Auerbach et al.}, 2001].
Derived from the civil-engineering methods developed for
the safety of human-built structures, including dams and bridges, the
standard approach to slope instability is to identify
the conditions under which a slope becomes unstable
[e.g. {\it Hoek and Bray}, 1997]. In this class of approach, geomechanical
data and properties are inserted in finite elements or discrete elements
numerical codes to predict the  possible departure from static equilibrium
or the distance to a failure threshold. The results of such analyses are
expressed using a safety  factor $F$, defined as the ratio between the
maximum retaining force to the driving forces.
According to this approach, a slope becomes unstable when $F<1$.
This approach is at the basis of landslide hazard maps, 
using safety factor value $F$ larger than $1$.

By their nature, standard stability analysis
does not account for
acceleration in slope movement [e.g. {\it Hoek and Brown}, 1980].
The problem is
that this modeling strategy gives a nothing-or-all signal.
In this view, any specific landslide is essentially unpredictable,
and the focus is on the recognition of landslide prone areas.
Other studies of landslides analyze the propagation
of a landslide and try to predict the maximum runout length of a landslide
[{\it Heim}, 1932, {\it Campbell}, 1989; 1990].
These studies do not describe the initiation of a catastrophic collapse.
To account for a progressive slope failure, i.e., a time
dependence in stability analysis, previous work have taken a quasi-static
approach in which some parameters are taken to vary slowly to account for
progressive changes of external conditions and/or external loading.
For instance, the accelerated motions have been linked to pore
pressure changes [{\it e.g. Vangenuchten and Derijke},
1989;  {\it Van Asch et al.}, 1999]. According to this approach,
an instability occurs when the gravitational pull on a slope becomes
larger than the
resistance of a particular subsurface level. This resistance on a
subsurface level is controlled by the friction coefficient of the
interacting surfaces. Since pore pressure acts at the level of
submicroscopic to macroscopic discontinuities, which themselves
control the global friction coefficient,
circulating water can hasten chemical alteration of the
interface roughness, and pore pressure itself can force adjacent
surfaces apart [{\it Vangenuchten and Derijke}, 1989].
Both effects can lead to a reduction in the friction
coefficient that leads, when constant loading applies, to accelerating
movement. However, this explanation has not yielded
quantitative method for forecasting slope movement.

Other studies proposed that (i) rates of slope movements are controlled
by microscopic slow cracking, and (ii) when a major failure plane is
developed, the abrupt decrease in shear resistance may provide a
sufficiently large force imbalance to trigger a catastrophic slope rupture
[{\it Kilburn and Petley}, 2003]. Such a mechanism, with a proper law
of input of
new cracks, may reproduce the acceleration preceding the collapse
that occurred at Vaiont, Mt Toc, Italy [{\it Kilburn and Petley}, 2003].

An alternative modeling strategy consists
in viewing the accelerating displacement of the slope prior to the
collapse as the final stage of the tertiary creep preceding
failure [{\it Saito and Uezawa}, 1961; {\it Saito}, 1965, 1969;
{\it Kennedy and Niermeyert}, 1971; {\it Kilburn and Petley}, 2003].
Further progress in exploring the relevance of this mechanism
requires a reasonable knowledge of the geology of the sliding surfaces,
their stress-strain history, the mode of failure, the time-dependent shear
strength and the piezometric water level values along the surface of failure
[{\it Bhandari}, 1988]. Unfortunately, this information is hard
to obtain and usually not available.
This mechanism,
viewing the accelerating displacement of the slope prior to the
collapse as the final stage of the tertiary creep preceding
failure, is therefore used mainly as a justification for
the establishment of empirical criteria of impending landslide
instability. Controlled experiments on landslides driven by a monotonic
load increase at laboratory scale have been quantified by a scaling law relating
the surface acceleration $d\dot{\delta}/dt$
to the surface velocity $\dot{\delta}$ according to
\be
d\dot{\delta}/dt = A \dot{\delta}^{\alpha}~,
\label{mgnlsa}
\ee
where $A$ and ${\alpha}$ are
empirical constants [{\it Fukozono}, 1985]. For ${\alpha}>1$, this relationship
predicts a divergence of the sliding velocity in finite time at some critical
time $t_c$. The divergence is of course not to be taken literally: it signals
a bifurcation from accelerated creep to complete slope instability
for which inertia is no more negligible. Several cases have been
quantified ex-post with this law, usually for ${\alpha}=2$, by plotting the
time $t_c-t$ to failure as a function of the inverse of the creep velocity
(see for a review [{\it Bhandari}, 1988]). Indeed, integrating
(\ref{mgnlsa}) gives
\be
t_c-t \sim \left({1 \over \dot{\delta}}\right)^{1 \over \alpha-1}~.
\label{jmgjkla}
\ee
These fits suggest that it might be possible to forecast
impending landslides by recording accelerated precursory slope
displacements. Indeed, for the Mont Toc, Vaiont landslide revisited here,
{\it Voight} [1988] mentioned that a prediction of the failure date
could have been made more than 10 days before the actual failure,
by using a linear relation linking the inverse velocity and the time
to failure, as found from (\ref{jmgjkla}) for $\alpha=2$. Our goal will
be to avoid such an a priori postulate by calibrating a more general
physically-based model.
{\it Voight} [1988, 1989] proposed that the relation (\ref{mgnlsa}),
which generalizes damage mechanics laws [{\it Rabotnov}, 1969; {\it Gluzman
and Sornette}, 2001], can be used with other variables (including
strain and/or seismic energy release) for a large variety of materials
and loading conditions. Expression (\ref{mgnlsa})
seems to apply as well to diverse types of landslides occurring in
rock and soil, including first-time and reactivated slides [{\it
Voight}, 1988].
It may be seen as a special case of a general expression for failure
[{\it Voight}, 1988, 1989].
Recently, such time-to-failure laws have been interpreted
as resulting from cooperative critical phenomena and
have been applied to the prediction of failure of heterogeneous composite
  materials [{\it Anifrani et al.}, 1995] and to precursory increase of seismic
  activity prior to main shocks [{\it Sornette and Sammis}, 1995;
{\it Jaume and Sykes}, 1999; {\it Sammis and Sornette},
  2002]. See also [{\it Sornette}, 2002] for extensions to other fields.

Here, we focus on two case studies, La Clapi\`ere sliding system in
the French Alps and the Vaiont landslide in the Italian Alps.
The latter landslide led to a  catastrophic collapse after 70 days
of recorded velocity increase. In the former case study, decades of
accelerating
motion aborted and gave way to a slow down of the system.
First, we should stress that, as for earthquakes for instance,
it is extremely difficult to obtain all relevant geophysical parameters
that may be germane to a given landslide instability.
Furthermore, it is also a delicate exercise to scale up the
results and insights obtained from experiments performed in the laboratory
to the scale of mountain slopes. Having said that, probably the simplest
model of landslides considers the moving part of the landslide
as a block sliding over a surface endowed with some given topography.
Within such a conceptual model, the complexity of the
landsliding behavior emerges from (i) the dynamics of the block behavior
(ii) the dynamics of interactions between the block and the substratum,
(iii) the history of the external loading (e.g. rain, earthquake).
In the following, we develop a simple model of sliding instability based
on rate and state dependent solid friction laws and we test how the 
friction law of a rigid block driven by constant gravity force can be 
useful for understanding the apparent transition between slow stable 
sliding and fast unstable sliding leading to slope collapse.

Previous modeling efforts of landslides in terms of a rigid
slider-block have taken
either a constant friction coefficient or a slip- or velocity-dependent
friction coefficient between the rigid block and the surface.
A constant solid friction coefficient (Mohr-Coulomb law)
is often taken to simulate bed- over bed-rock sliding.
{\it Heim} [1932] proposed this model as an attempt to predict the propagation
  length of rock avalanches. In this pioneering study
to forecast extreme runout length, the constant friction coefficient was
interpreted as an effective average friction coefficient.
In contrast, a slip-dependent friction coefficient model is taken to simulate
the yield-plastic behavior of a brittle material beyond
the maximum of its strain-stress characteristics.
For rock avalanches, {\it Eisbacher} [1979] suggested that
the evolution from a static to a dynamic friction coefficient
is induced by the emergence of a basal gouge.
Studies using a velocity-dependent friction coefficient have mostly
focused on the establishment of empirical relationships
between shear stress $\tau$ and block velocity $v$,
such as $v \sim \exp (a \tau)$ [{\it Davis et al.}, 1990] or
$v \sim \tau^{1/2}$ [{\it Korner}, 1976], with however no
definite understanding of the possible mechanism
[see for instance  {\it Durville}, 1992].

Our approach is to account
for the interaction between the block and the underlying slope by a solid
friction law encompassing both state and velocity dependence, as established
by numerous laboratory experiments (see for instance [{\it Scholz}, 1990,
1998; {\it Marone}, 1998; {\it Gomberg et al}., 2000] for reviews).
The sliding velocities used in laboratory to establish the rate and state
friction laws are of the same order,  $10^{-4}-10^2$ $\mu$m/s,  than those
observed for landslides before the catastrophic collapse.
On the one hand, state- and velocity-dependent
friction laws have been developed and used extensively to model
the preparatory as well as the elasto-dynamical phases of earthquakes.
On the other hand, analogies between landslide faults and tectonic faults
have been noted [{\it Gomberg et al.}, 1995] and the use of the static
friction coefficient is ubiquitous in the analysis of slope stability.
However, to our knowledge, no one has pushed further the analogy
between sliding rupture and earthquakes and no one has used the physics
of state- and velocity-dependent friction to apply it to the problem of
landslides and their precursory phases.
Such standard friction laws have been shown to lead to an asymptotic
time-to-failure power law with $\alpha=2$ in the late stage of
frictional sliding
motion between two solid surfaces preceding the elasto-dynamic rupture
instability [{\it Dieterich}, 1992].
This model therefore accounts for the finite-time singularity of the sliding
velocity (\ref{jmgjkla}) observed for landslides and rationalizes the empirical
time-to-failure laws proposed by {\it Voight} [1988, 1990].
In addition, this model also describes the stable
sliding regime, the situation where the time-to-failure behavior is absent.

In the first section, we derive the four different sliding regimes of
this model which depend on the ratio $B/A$ of two parameters of
the rate and state friction law and on the initial conditions of
the reduced state variable. Sections 3 and 4 analyze the Vaiont and
 La Clapi\`ere landslides, respectively.
In particular, we calibrate the slider-block model
to the two landslide slip data and invert the key parameters. Of particular
interest is the possibility of distinguishing between an unstable
and a stable sliding regime. Our results suggest the Vaiont landslide
(respectively La Clapi\`ere landslide) as belonging to the velocity weakening
unstable (respectively strengthening stable) sliding regime.
Section 5 concludes. A companion paper investigates the potential of
our present results for landslide prediction: the predictability of
the failure times and prediction horizons are investigated using
different methods [{\it Sornette et al.}, 2003].

\section{Slider-Block model with state and velocity dependent friction}

\subsection{Basic formulation}

Following [{\it Heim}, 1932; {\it Korner}, 1976; {\it Eisbacher}, 1979;
{\it Davis et al.}, 1990; {\it Durville}, 1992], we model the future
landslide as a block resting on an inclined slope forming an angle
$\phi$ with respect to the horizontal.
In general, the solid friction coefficient $\mu$ between two surfaces
is a function of the cumulative slip $\delta$ and of the slip velocity
$\dot{\delta}$.
There are several forms of rate/state-variable constitutive law that have
been used to model laboratory observations of solid friction. The version
currently in best agreement with experimental data, known as the
Dieterich-Ruina or `slowness' law  [{\it Dieterich}, 1978; {\it Ruina}, 1983],
is expressed as
\be
\mu = \mu_0 + A \ln {\dot{\delta} \over \dot{\delta}_0} + B \ln
{\theta \over \theta_0}~,
\label{vxcxxzx}
\ee
where the state variable $\theta$ is usually interpreted as
proportional to the surface
of contact between asperities of the two surfaces. $\mu_0$ is the
friction coefficient for a sliding velocity $\dot{\delta}_0$
and a state variable $\theta_0$.
The state variable $\theta$  evolves with time according to
\be
{{\rm d}\theta \over {\rm d}t} = 1 - {\theta \dot{\delta} \over
D_c}~, \label{qwertyu}
\ee
where $D_c$ is a characteristic slip distance, usually interpreted as
the typical size of asperities. Expression (\ref{qwertyu})
can be rewritten as
\be
{{\rm d}\theta \over {\rm d}\delta} = {1 \over \dot{\delta}}
- {\theta \over D_c}~. \label{qweadasgatyu}
\ee

As reviewed in [{\it Scholz}, 1998],
the friction at steady state is:
\be
\mu_S = {\hat \mu}_0 + (A-B) \ln {\dot{\delta} \over \dot{\delta}_0}~,
\ee
where ${\hat \mu}_0 = \mu_0 + B \ln {D_c \over \theta_0 \dot{\delta}_0}$.
Thus, the derivative of the steady-state friction coefficient with
respect to the logarithm of the reduced slip velocity is $A-B$. If
$A>B$, this derivative is positive: friction increases with slip velocity
and the system is stable as more resistance occurs which tends to react against
the increasing velocity. In contrast, for $A<B$, friction exhibits the
phenomenon of velocity-weakening and is unstable.

The primary parameter that determines stability, $A-B$, is a
material property. For instance, for granite, $A-B$ is negative
at low temperatures and becomes positive for temperatures above
about 300$^o$ C. In general, for low-porosity
crystalline rocks, the transition from negative to positive $A-B$
corresponds to a change from elastic-brittle deformation to crystal
plasticity in the micro-mechanics of friction [{\it Scholz}, 1998].
For the application to landslides, we should in addition consider that
sliding surfaces are not only contacts of bare rock surfaces: they
are usually lined with wear detritus, called cataclastic or fault gouge.
The shearing of such granular material involves an additional
hardening mechanism (involving dilatancy), which tends to make
$A-B$ more positive. For such materials, $A-B$ is positive when
the material is poorly consolidated, but decreases at elevated
pressure and temperature as the material becomes lithified.
See also section 2.4 of Scholz's book [{\it Scholz}, 1990].

The friction law (\ref{vxcxxzx}) with (\ref{qwertyu}) accounts for
the fundamental properties of a broad range of surfaces in contact,
namely that they strengthen logarithmically when aging at rest,
and weaken (rejuvenate) when sliding [{\it Scholz}, 1998].

To make explicit the proposed model, let us represent schematically
a mountain flank as a system made of a block and
of its basal surface in which it is encased.
The block represents the part of the slope which may be potentially
unstable. For a constant gravity loading, the two parameters
controlling the stability of the block are the dip angle $\phi$
between the surface on which the block stands and the horizontal
  and the solid friction coefficient $\mu$.
  The block exerts stresses that are normal ($\sigma$)
as well as tangential ($\tau$) to this surface of contact.
The angle $\phi$ controls the ratio of the shear
over normal stress: $\tan \phi = \tau/\sigma$.
In a first step, we assume for simplicity that the usual
solid friction law $\tau = \mu \sigma$ holds for all times,
expressing that the shear stress $\tau$ exerted on the block
is proportional to the normal stress with a coefficient of
proportionality defining the friction coefficient $\mu$. This assumption
expresses a constant geometry of the block and of the surface of sliding.
For the two landslides that we study in this paper, a rigid block
sliding on a slope with a constant dip angle is a good first order approximate
of these landslide behaviors.

\subsection{Solution of the dynamical equation}

\subsubsection{Asymptotic power law regime for $A-B<0$}

As the sliding accelerates, the sliding velocity becomes sufficiently large
such that $\dot{\delta} \gg D_c/\theta$ and we can neglect the first term
  $1/\dot{\delta}$ in the right-hand-side of (\ref{qweadasgatyu})
[{\it Dieterich}, 1992]. This yields
\be
\theta = \theta_0 \exp\left(-\delta/D_c\right)~,   \label{njjfa}
\ee
which means that $\theta$ evolves toward zero. The friction law then reads
\be
{\tau \over \sigma} =
\mu_0 + A \ln {\dot{\delta}\over \dot{\delta}_0} - {B \delta \over
D_c}~,  \label{fjjanffg}
\ee
where we have inserted (\ref{njjfa}) into (\ref{vxcxxzx}).
In this equation, $\tau$ and $\sigma$ result from the mass of the block
and are constant. The solution of (\ref{fjjanffg}) is [{\it Dieterich}, 1992]
\be
\delta (t) = - {A D_c \over B}~~ \ln\left[{B \dot{\delta}_0
~e^{{\tau \over \sigma}-\mu_0 \over A} \over A D_c}~(t_c -t)\right]~,
\label{nglnwq}
\ee
where $t_c$ is determined by the initial condition $\delta(t=0)
\equiv \delta_i$ :
\be
t_c={A D_c \over B \dot{\delta}_0}
~ e^{-\left({B \delta_i\over A D_c} + {{\tau \over \sigma}-\mu_0
\over A}\right)}
\label{tc}
\ee
The logarithmic blow up of the cumulative slip in finite time is associated
with the divergence of the slip velocity
\be
\dot{\delta} = {A D_c \over B}~~ {1 \over t_c -t}~,
\label{ngaaz}
\ee
which recovers (\ref{jmgjkla}) for $\alpha=2$.

\subsubsection{The complete solution for the frictional problem}

The solution (\ref{nglnwq}) is valid only for $A-B<0$ and sufficiently
close to $t_c$ for which the slip velocity $\dot{\delta}$ is large,
ensuring the validity of the approximation leading to (\ref{njjfa}).
However, even in the unstable case $A-B<0$, the initiation of sliding
cannot be described by using the approximation established for $t$
close to $t_c$ and
requires a description different from (\ref{nglnwq}) and (\ref{ngaaz}).
Furthermore,
we are interested in different situations, in which the sliding may not
always result into a catastrophic instability, as for instance for
the mountain slope La Clapi\`ere, which started to slip but did not reach
the full instability, a situation which can be interpreted
as the stable regime $A-B>0$.
The complete solution for the frictional problem is derived in Appendix A.

\subsection{Synthesis of the different slipping  regimes}

\begin{figure}
\psfig{file= 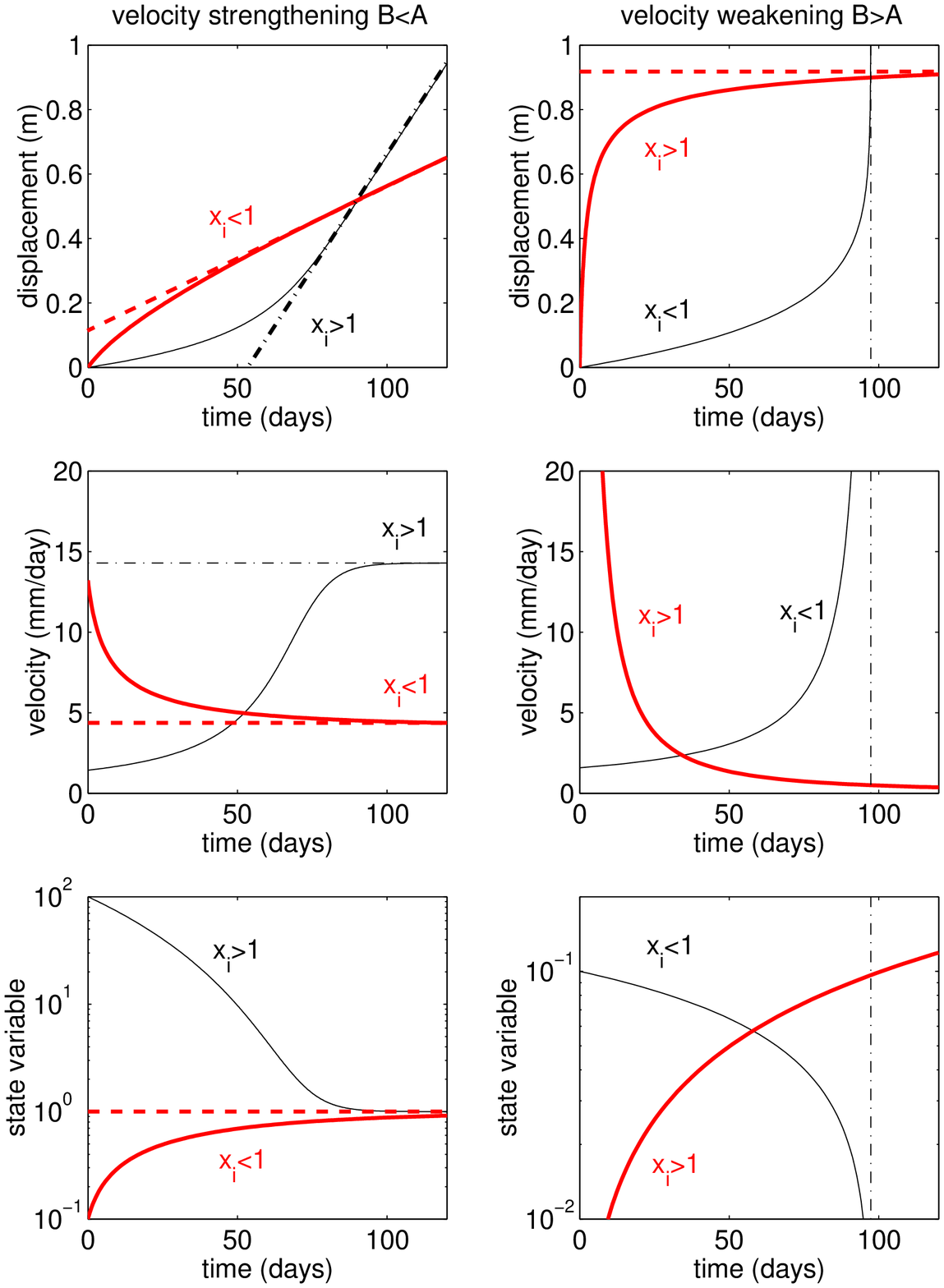,width=11cm}
\caption{\label{classregime}  Schematic classification of the different
regimes of sliding discussed in the text. The left column
of three panels correspond to the stable regime $m=B/A<1$
and the right column of three panels describes the unstable
regime $m=B/A>1$. In each case, the displacement, velocity
and state variables are shown as a function of time.
Each regime (stable and unstable) is divided into two
cases, depending on the dimensionless initial value
$x_i \propto \theta_i$ of the state
variable. The thick lines corresponds to decreasing velocities
and increasing state variables. The thin lines correspond to
increasing velocities and decreasing state variables.}
\end{figure}

The block sliding displays different regimes as a function
of the friction law parameters and of the initial conditions.
These regimes are controlled by the value of the friction law
parameters, i.e., the parameter $m=B/A$, the initial value
$x_i$ of the reduced state variable and the material parameter $S$
defined by (\ref{mmjdl}).
$A$ and $B$ are defined in (\ref{vxcxxzx}) and are determined
by material properties. $x_i$ is  defined in (\ref{mgmbm})
and is proportional to the initial value of the
state variable $\theta$. The parameter $S$ is independent
of the initial conditions. As derived from the complete solution
in Appendix A, the different regimes are summarized below and
in Table 1 and illustrated in  Figure  \ref{classregime}.
\begin{itemize}
\item
For $0<m<1$ the sliding is always stable.
Depending of the initial value for $t=0$ of the reduced state variable $x_i$,
the sliding velocity either increases (if $x_i>1$) or decreases (if $x_i<1$)
toward a constant value.
\item
For $m>1$ the sliding is always unstable.
When  $x_i<1$, the sliding
velocity  increases toward a finite-time singularity.
The slip velocity diverges as $1/(t_c-t)$ corresponding
to a logarithmic singularity of the cumulative slip.
For $x_i>1$,  the velocity decreases toward a vanishingly small value.
\end{itemize}

  \subsection{Analysis of landslide observations, {\it, applications to landslide behaviors}}
In the sequel, we test how this model can reproduce the observed acceleration
of the displacement for Vaiont and La Clapi\`ere landslides.
The Vaiont landslide was the catastrophic culmination of an accelerated slope
velocity over a two months period [{\it Muller}, 1964].
La Clapi\`ere landslide was characterized by a long lasting acceleration that peaked up
in the 1986-1988 period, succeeded by a restabilizing phase
[{\it Susella and Zanolini}, 1996].
An acceleration of the displacement can arise from the friction model in
two regimes, either in the stable regime with $m<1$ and $x_i>1$ or in the
unstable regime with $m>1$ and $x_i<1$. In the first case, the acceleration
evolves toward a stable sliding. In the unstable case, the acceleration leads
to a finite-time singularity of the displacement and of the velocity.
However, these two regimes are very similar in the early time regime
before the critical time (see Figure  \ref{classregime}). It is therefore very
difficult to distinguish from limited observations a landslide in the stable
regime from a landslide in the unstable regime when far from the rupture.

We assume that the friction law parameters, the geometry of the landslide
and the gravity forces are constant. Within this conceptual model, the
complexity of the landsliding  behavior emerges from the friction law.
We are aware of neglecting in this  first order analysis any possible
complexity inherent either to the geometry and rheology of a larger set
of blocks, or the geometry and rheology of the substratum or the history
of the external loading (e.g. earthquake, rainfalls).
We invert the friction law parameters from the velocity and displacement data
of the Vaiont and La Clapi\`ere landslides.
Our goal is (i) to test if this model is useful for distinguishing an unstable
accelerating sliding characterized by $B>A$ from a stable accelerating regime
occurring for $B<A$ and
(ii) to test the predictive skills of this model and compare with other
methods of prediction.

\section{The Vaiont landslide}
\subsection{Historical and geo-mechanical overview}

On the Mt Toc slope in the Dolomite region in the Italian Alps
about 100 km north of Venice,
on October 9, 1963, a 2 km-wide landslide was initiated at an elevation of
1100-1200 m, that is 500-600 m above the valley floor. The event ended
70 days later in a 20 m/s run-away of
about 0.3 km$^3$ of rocks sliding into a dam reservoir.
The high velocity of the slide triggered  a water surge within the reservoir,
overtopping the dam and killing 2500 people in the villages
(Longarone, Pirago, Villanova, Rivalta and Fae) downstream.

This landslide has a rather complex history. The landslide occured on
the mountain above a newly built dam reservoir.
The first attempt to fill up the reservoir was made between March
and November 1960. It induced recurrent observations of creeping motions
of a large mass of rock above the reservoir, and led to several small and
rather slow slides [{\it Muller}, 1964]. Lowering the
reservoir water level induced the rock mass velocities to drop from
$\sim 40$ mm/day to $< 1$ mm/day. A controlled raising of
the water level as well as cycling
were performed.
A second peak of creeping velocity, at about 10 mm/day
was induced by the 1962 filling cycle.
The 1963 filling  cycle started in April. From May, recurrent
increases of the creep velocity were measured using 4 benchmarks.
On september 26, 1963,
lowering the reservoir level was again initiated. Contrary to what happened in 1960 and
again in 1962, the velocities continued to increase at an increasing rate. This culminated
in the $20$ m/s downward movement of a volume of 0.3 km$^3$ of rock
in the reservoir.

The landslide geometry is a rough rectangular shape, 2 km wide and
1.3 km in length.
Velocity measurements are available for four benchmarks, corresponding to
four different positions on the mountain slope, respectively denoted 5, 50,
63 and 67 in the Vaiont nomenclature. Benchmarks 63 and 67 are located at the
same elevation in the upper part of the landslide a few hundred meters from the
submittal scarp. The distance between the two benchmarks is 1.1 km.
The benchmark 5 and 50 are 700 m downward the 63-67 benchmark level.

\begin{figure}
\psfig{file=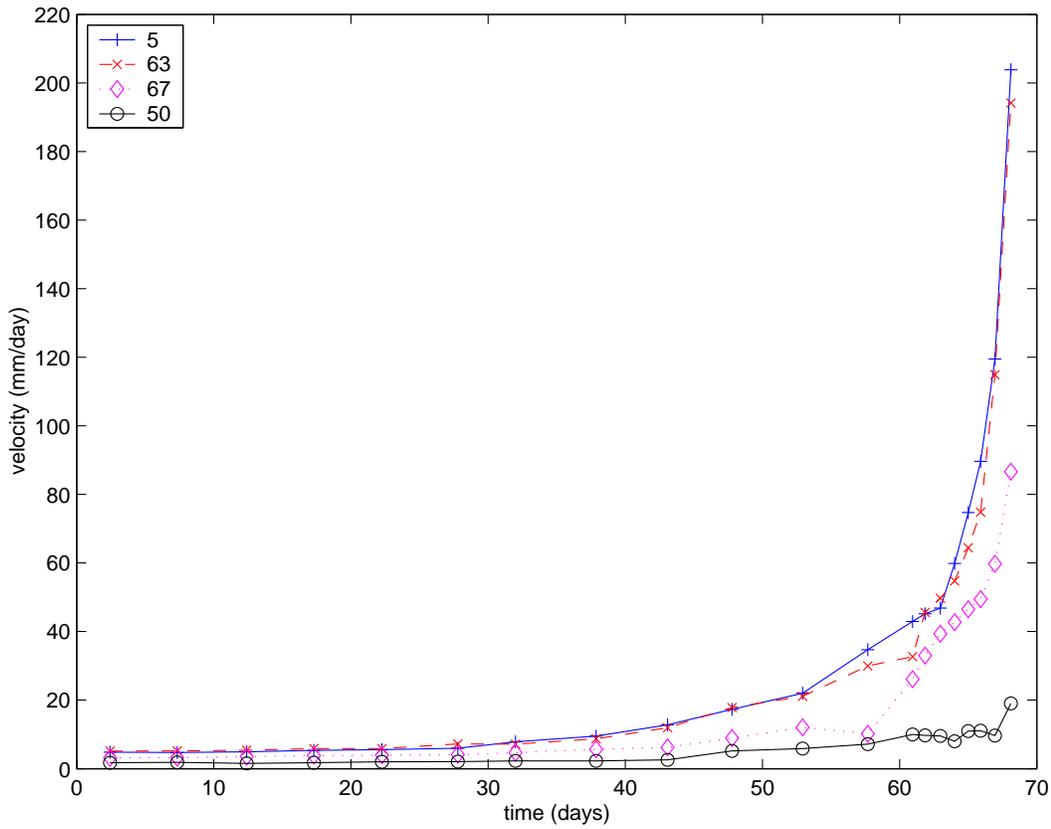,width=14cm}
\caption{\label{olkdz} Velocity measurements for the four benchmarks
of the Vaiont landslide. Benchmarks 5 and 63 exhibit similar
  acceleration. Benchmark 50
shows only a relatively small acceleration in absolute values at the end of
the 60 days accelerating phase. Its acceleration is however significant
  in relative values, as seen in Figure \ref{vvaiontv}. Data from 
[{\it Muller}, 1964].}
\end{figure}

Figure \ref{olkdz} shows the velocity of the four benchmarks on the block as a
function of time prior to the Vaiont landslide.
For these four benchmarks, the deformation of the sliding zone prior to
rupture is not homogeneous, as the cumulative displacement in the period
from August 2$^{nd}$, 1963 to October 8, 1963 ranges from 0.8 to 4 m.
However, the low degree of disintegration of the distal
deposit [{\it Erismann and Abele}, 2000] argue for a possible homogeneous
block behavior during the 1963 sliding collapse.

It was recognized later that limestones and clay beds
dipping into the valley provide conditions favorable for dip-slope
failures [{\it Muller}, 1964, 1968; {\it Broili}, 1967].
There is now a general agreement on the collapse history of the 1963 Vaiont
landslide (see e.g., [{\it Erismann and Abele}, 2000]). The failure occurred
along bands of clays within the limestone mass at depths between 100-200 m
below the surface [{\it Hendron and Patton}, 1985]. Raising the reservoir
level increased water pore pressure in the slope flank, that triggered
failure in the
clays layers. Final sliding occurred after 70 days of down-slope
accelerating movement. The rock mass velocity progressively
increased from 5 mm/day to more than 20 cm/day, corresponding to a
cumulative displacement of a few meters over this 70 days period
[{\it Muller}, 1964].

\subsection{Analysis of the velocity data with the
slider-block model parameters.}

\begin{figure}
\psfig{file=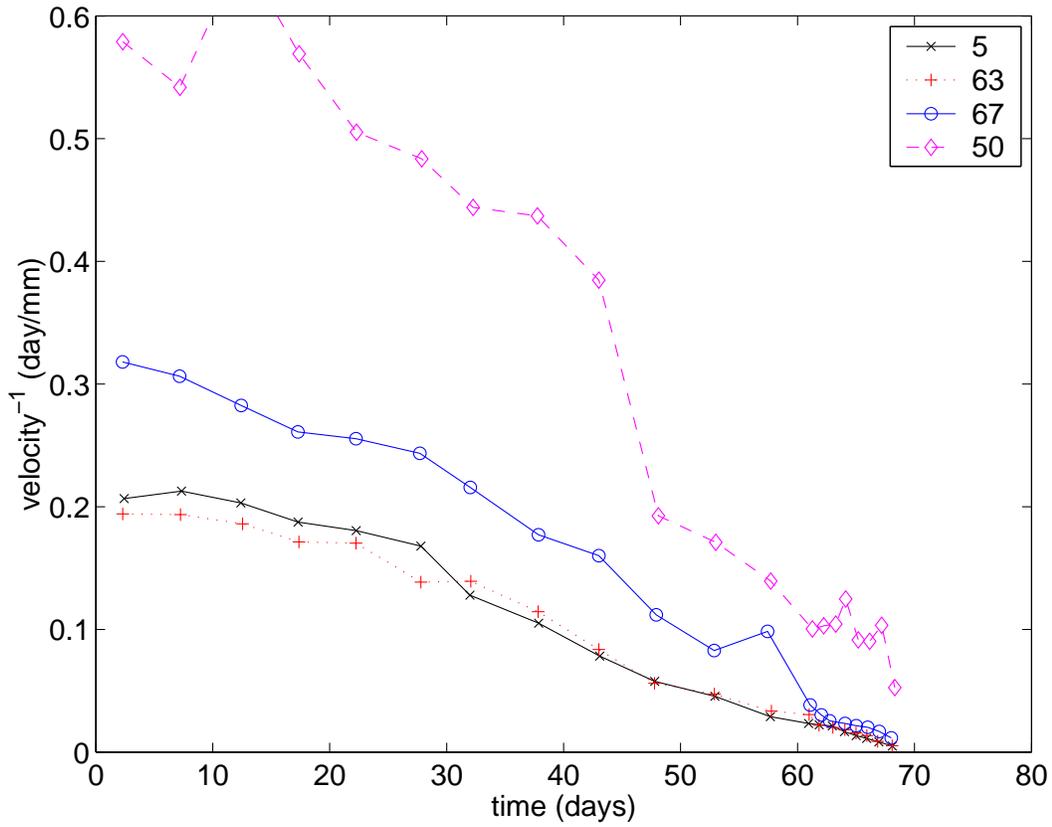,width=14cm}
\caption{\label{olkdzz} Same as Figure \ref{olkdz} by plotting the inverse
of the velocity  as a time $t$.
All curves are roughly linear, showing that the velocity
exhibits a finite-time singularity $v \sim 1/(t_c-t)$ with $t_c 
\approx 69.5$ days
for all benchmarks, estimated as the intercept of the extrapolation of these
curves with the horizontal axis.}
\end{figure}

Figure \ref{olkdzz} shows the inverse of the velocity shown in Figure
\ref{olkdz} to test the finite-time-singularity hypothesis
(\ref{jmgjkla},\ref{ngaaz}).
Note that this figure does not require the knowledge of the
critical time $t_c$ and is not a fit to the data.
The curves for all benchmarks are roughly linear in this representation,
in agreement with a finite-time singularity of the velocity (\ref{jmgjkla})
with $\alpha=2$. It was the observations presented in Figure \ref{olkdzz} that
led Voight to suggest
that a prediction could have been issued more than 10 days before the collapse
[{\it Voight}, 1988].
We note that the law $\dot{\delta} \propto 1/(t_c-t)$ requires the
adjustment of
$\alpha$ to the special value $2$ in the phenomenological approach
[{\it Voight}, 1988] underlying (\ref{jmgjkla}) while it is a robust and
universal result in our model leading to (\ref{ngaaz}) in the
velocity-weakening
regime $B>A$, $m>1$ and for a normalized initial state variable larger than $1$
(see equation (\ref{ngaaz}) and Table 1).

In order to invert the parameters $m$, $D$, $T$ of the friction model
and the initial condition of the state variable $x_i$ from the
velocity data, we minimize the $rms$ (root-mean-square)
of the residual between the observed velocity $\dot{\delta}_{obs}$
and the velocity $\dot{\delta}$ from the friction model
(\ref{ode}) and (\ref{velocity}).
The constant $D$ in (\ref{velocity}) is obtained by taking the
derivative of the $rms$ with respect to $D$, which yields
\be
    D={ \sum_{t_i} \dot{\delta}(t_i) \dot{\delta}_{obs}(t_i)
   \over \sum_{t_i} {\dot{\delta}(t_i)}^2}
\label{D}
\ee
where the  velocity  $\dot{\delta}$ in (\ref{D}) is evaluated for $D=1$
in (\ref{velocity}).
We use a simplex algorithm (matlab subroutine) to invert the three other
parameters. For each data set, we use different starting points (initial
parameter values for the simplex algorithm) in the inversion to test
for the sensitivity of the results on the starting point.

\begin{figure}
\psfig{file=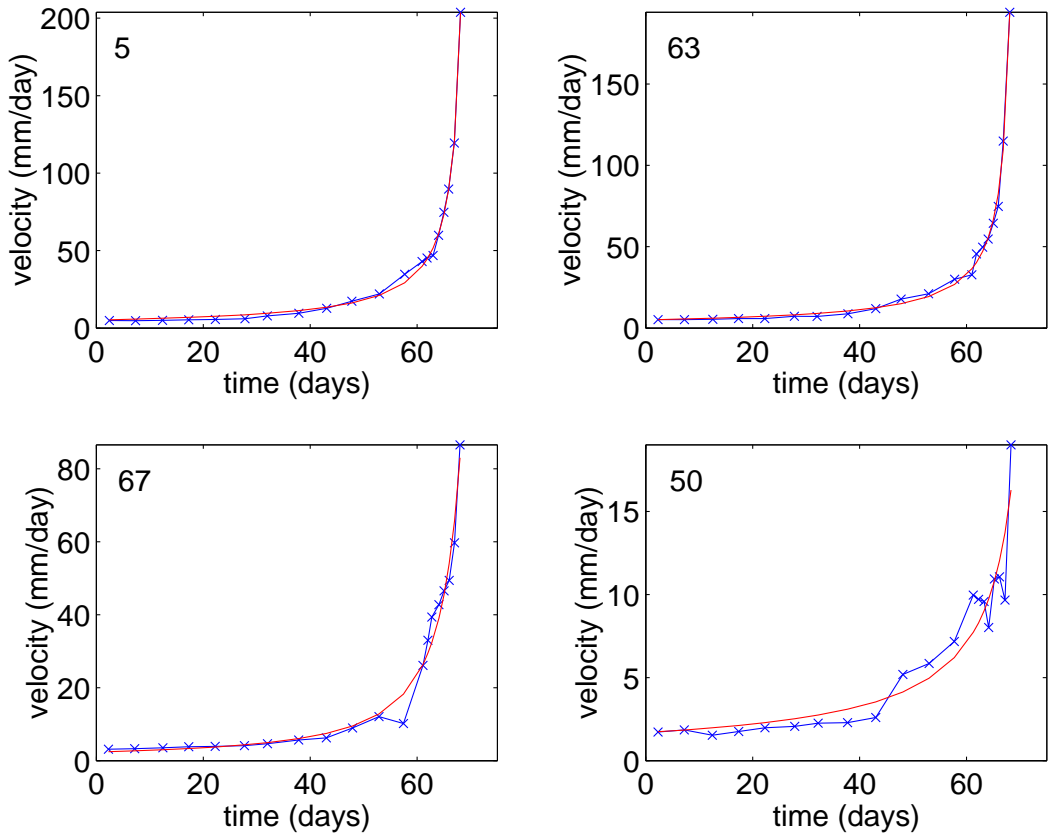 ,width=14cm}
\caption{\label{vvaiontv} For each of the four Vaiont benchmarks,
the velocity data of Figure \ref{olkdz}
is fitted with the slider-block model with the state and velocity
friction law (\ref{ode}) and (\ref{velocity}) by adjusting the set of
parameters $m$, $D$, $T$ and the initial condition of the state variable $x_i$.
The data are shown as the crosses linked by straight segments and the fit
is the thin continuous line. The fitted $m$ are respectively
$m=1.35$ (benchmark 5), $m=1.24$ (benchmark 63), $m=0.99$ (benchmark 67)
and $m=1.00$ (benchmark 50).}
\end{figure}

Figure \ref{vvaiontv} shows the fits to the velocity data
using the slider-block model with the state and velocity
friction law (\ref{velocity}) and (\ref{ode}). The values of $m=B/A$ are
respectively $m=1.35$ (benchmark 5), $m=1.24$ (benchmark 63), $m=0.99$
(benchmark 67) and $m=1.00$ (benchmark 50).
Most values are  larger than or equal to $1$, which is compatible with the
finite-time-singularity regime summarized in Table 1.
The parameters of the friction law are very poorly constrained by the
inversion.
In particular, even for those benchmarks were the best fit gives $m>1$,
other models with $m<1$ provide a good fit to the velocity with only slightly
larger rms.

\begin{figure}
\psfig{file=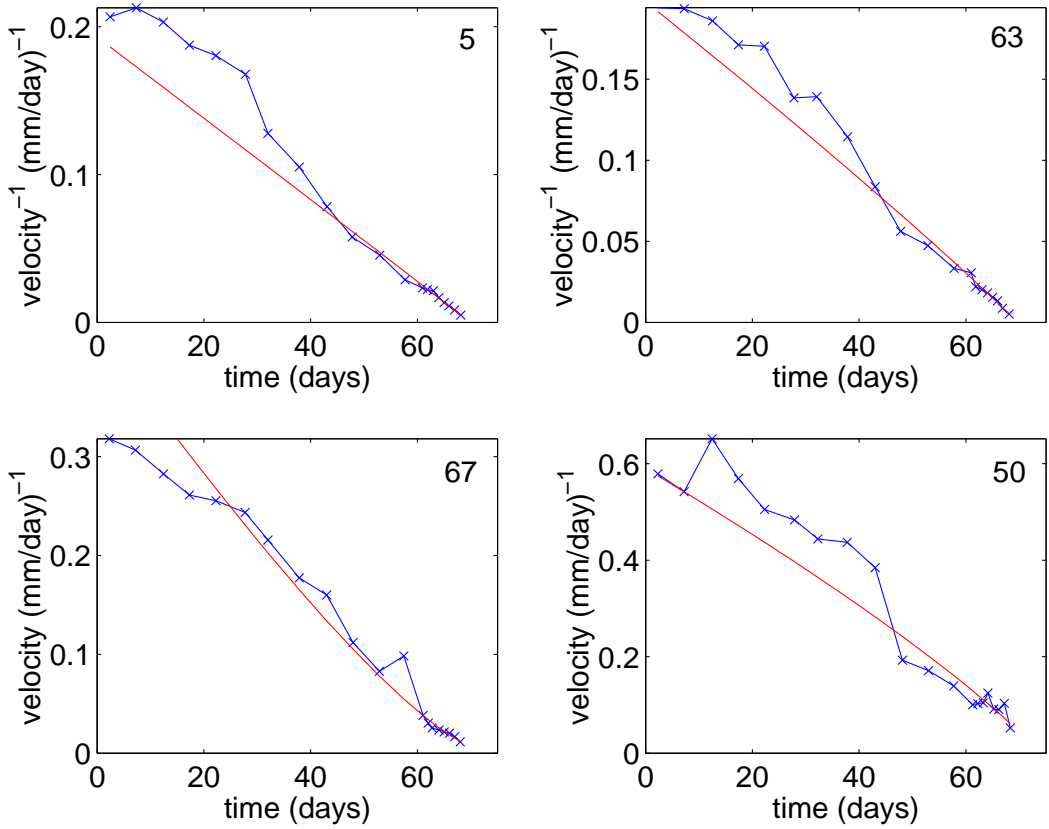 ,width=14cm}
\caption{\label{vvaiontvm1}
Same as Figure \ref{vvaiontv} but showing the inverse of the velocity.
The upward bending of the curve for benchmark 67 reflects the saturation
of the velocity in the stable regime $B<A$.
  The fit for the three other benchmarks characterized by $m \geq 1$ is very
  close to the  asymptotic solution $v \sim 1/(t_c-t)$ (\ref{ngaaz}).}
\end{figure}

Figure \ref{vvaiontvm1} gives another representation of Figure \ref{vvaiontv}
showing the inverse of the velocity as a function of time.
The increase of velocity seems to be exhausted before the critical time
for all benchmarks, which may explain the values $m<1$ sometimes obtained by
the inversion.

\section{La Clapi\`ere landslide: the aborted 1982-1987 acceleration}

We now report  results on another case which exhibited a
transient acceleration
which did not result in a catastrophic failure but re-stabilized.
This example provides what is maybe an example of the
  $m<1$ stable slip regime, i.e. $B<A$, as interpreted
within the friction model.

\subsection{Historical and geo-mechanical overview}

  \subsubsection{Geo-mechanical setting and Displacement history: 1950-2000}

La Clapi\`ere landslide is located at an elevation between 1100 m and 1800 m
on a 3000m high slope. The landslide has a width of about 1000 m.
  Figure \ref{photo-clapieres} shows La Clapi\`ere landslide in
1979 before the acceleration of the displacement, and in 1999 after the end of
the crisis.
The volume of mostly gneiss rocks implied in the landslide
is estimated to be around $50 \times 10^6$ m$^3$.
At an elevation of about 1300 m, a 80 m thick bed provides a more massive
and relatively stronger level compared with
the rest of relatively weak and fractured gneiss. The two lithological entities
are characterized by a change in mica content which is associated with
a change of the peak strength and of the elastic modulus by a factor two
[{\it Follacci et al.}, 1990, 1993]. Geomorphological criteria allow one to
distinguish three distinct sub-entities within the landslide, NW,
Central and SW respectively [{\it Follacci et al.}, 1988].

\begin{figure}
\psfig{file=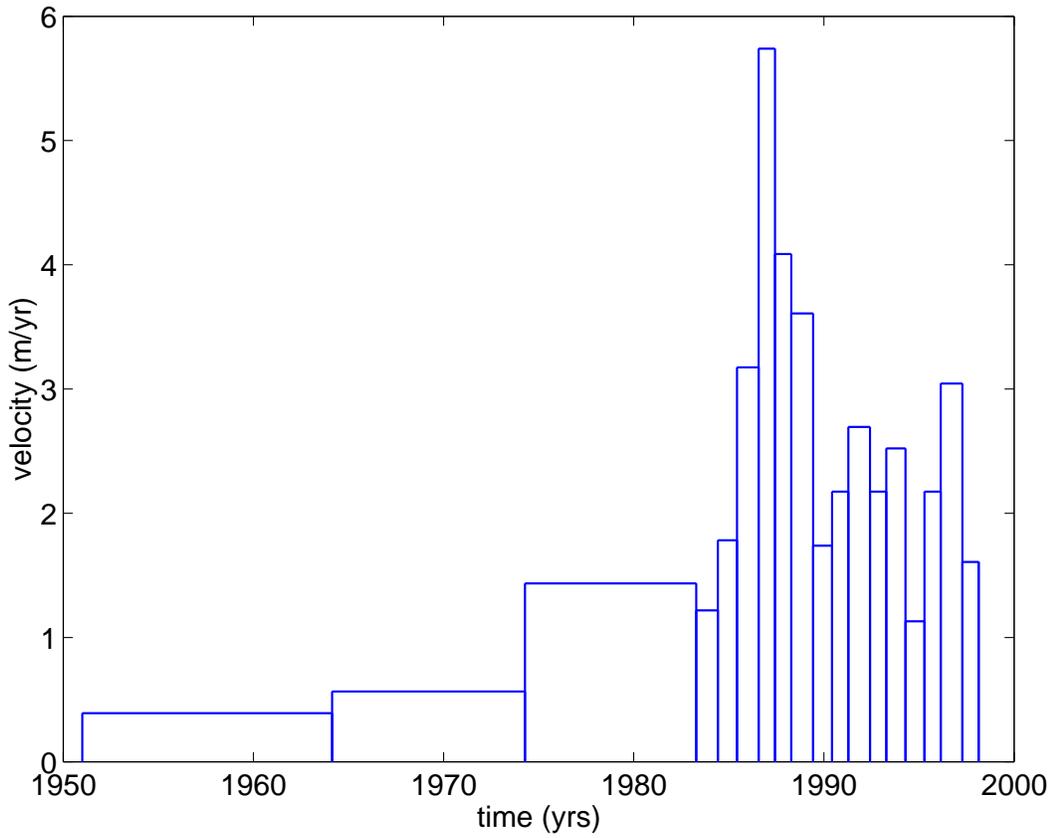,width=14cm}
\caption{ Velocity of the landslide of La Clapi\`ere
mount over almost 50 years, showing that the dangerous velocity peak in 1987
was preceded by a progressing build-up extending over several decades.
Before 1982, the velocity is inferred from aerial photographs in 
1951, 1964, 1974
and 1982. After 1982, the velocity is obtained from automated triangulation
and geodesy. Data from {\it CETE} [1999].}
\label{clapierelongterm}
\end{figure}

\begin{figure}
\psfig{file=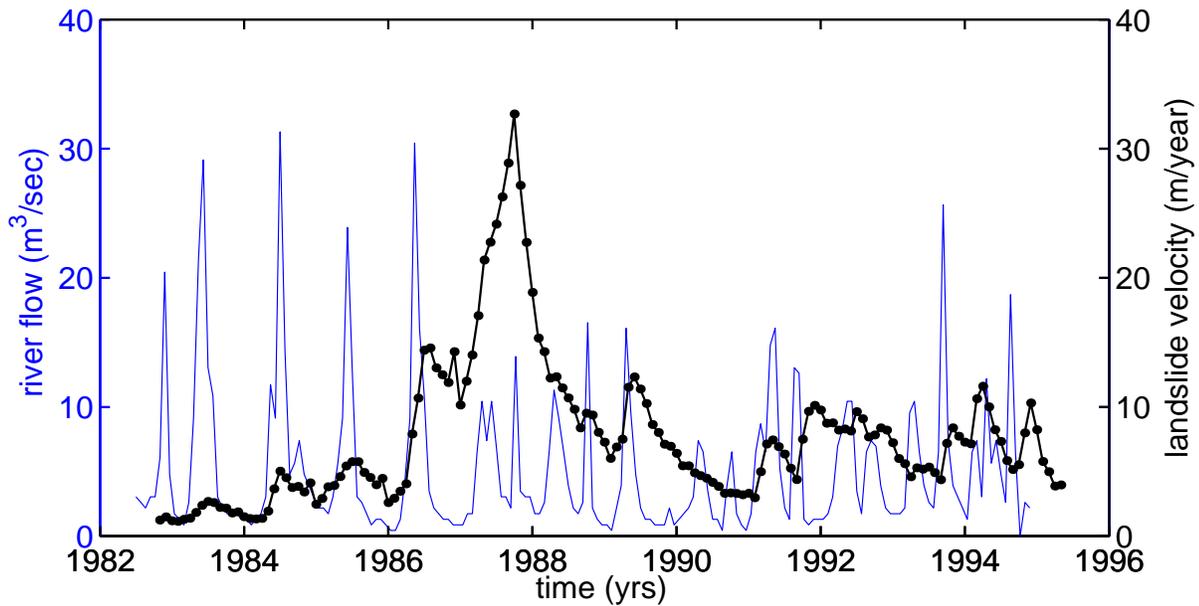,width=16cm}
\caption{Velocity pattern for benchmark 10 of  La Clapi\`ere 
landslide (solid line and dots)
and flow rates (thin solid line) of
the Tin\'ee river on the 1982-1995 period. Because the Tin\'ee river 
runs at the
basis of La Clapi\`ere landslide, the river flow rate reflects the
water flow within the landslide [{\it Follacci et al.}, 1993; {\it 
Susella and Zanolini}, 1996].
  The flow rates are measured at St Etienne village, 2 km upstream the 
landslide site.
There is no stream network on the landslide site. The Tin\'ee flow 
drains a 170 km$^2$ basin.
This tiny basin is homogeneous both in terms  of slopes and elevation 
(in the 1000-3000 m range).
Accordingly, the seasonal fluctuations of the river flow reflects
the amount of water within the landslide slope due to rainfalls and 
snow melting. Data from {\it CETE} [1996].
}
\label{debitpluie}
\end{figure}

There is some historical evidence that the rock mass started
to be active before the beginning of the 20th century. In 1938,
photographic documents attest the existence of a scarp at 1700 m elevation
[{\it Follacci}, 2000]. In the 1950-1980 period,
triangulation and aerial photogrametric surveys provide constraints on the
evolution of the geometry and the kinematics of the landslide
(Figure \ref {clapierelongterm}).
The displacement rate measured by aerial photogrametric survey increased from
0.5 m/yrs in the 1950-1960 period to 1.5 m/yrs in the 1975-1982 period
[{\it Follacci et al.}, 1988].
Starting in 1982, the displacements of 43 benchmarks have been monitored
on a monthly basis using distance meters (using
a motorised theodolite (TM300) and a Wild DI 3000 distance meter)
 [{\it Follacci et
al.}, 1988, 1993;
{\it Susella and Zanolini}, 1996]. The displacement data for the 5 benchmarks
  in Figure \ref{photo-clapieres} is shown in Figure \ref{4benmarkclap}.
The velocity of benchmark 10, which is typical, is shown in Figure
\ref{debitpluie}.
The rock mass velocities exhibited a dramatic increase between January 1986
and January 1988, that culminated in the 80 mm/day velocity  during
the 1987 summer and to 90 mm/day in October 1987.
The homogeneity of benchmark trajectories and the synchronous
acceleration phase
for most benchmark, attest of a global deep seated behavior of this landslide
[e.g. {\it Follacci et al.}, 1988].
However, a partitioning of deformation occurred, as reflected by the
difference in absolute values of benchmark displacements (Figure 
\ref{4benmarkclap}).
The upper part of the landslide moved slightly faster than the
lower part and the NW block.
The observed decrease in displacement rate since 1988 attest of a change
in landsliding regime at the end of 1987 (Figure \ref{4benmarkclap}) .

\subsubsection{Correlations between the landslide velocity and the river flow}

The landslide velocity displays large fluctuations correlated with fluctuations of
the river flow in the valley as shown in Figure \ref {debitpluie}.
There is a seasonal increase of the slope velocity which reaches a maximum
$V_{\rm max}$ of the order of or less than $30$ mm/days.
The slope velocity increases in the spring due to snow melting and over
a few days after heavy precipitations concentrated in the fall of each year
[{\it Follacci et al.}, 1988; {\it Susella and Zanolini}, 1996].
During the 1986-1988 period, the snow melt and rainfalls were not 
anomalously high
but the maximum value of the velocity, $V_{\rm max} = 90$ mm/day, was 
much larger
that the velocities reached during the 1982-1985 period for comparable
rainfalls and river flows [{\it Follacci et al.}, 1988; 1993].
This strongly suggests that the hydrological conditions are not the 
sole control
parameters explaining both the strong 1986-1987 accelerating and the 
equally strong
slowdown in 1988-1990.
During the interval 1988-1990, the monthly recorded velocities slowed down
to a level slightly higher than the pre-1986 values. Since 1988, the seasonal
variations of the average velocity never recovered the level established during
the 1982-1985 period [{\it Follacci et al.}, 1993; {\it David and ATM}, 2000].
{\it Rat} [1988] derives a relationship between the river flow and the
landslide velocity by adjusting an hydrological model to the velocity data
in the period 1982 to 1986. This model tuned to this time period
does not reproduce the observed acceleration of the velocity after 1986.

\subsubsection {Fracturing patterns contemporary to the 1986-1987
accelerating regime}

\begin{figure}
\psfig{file=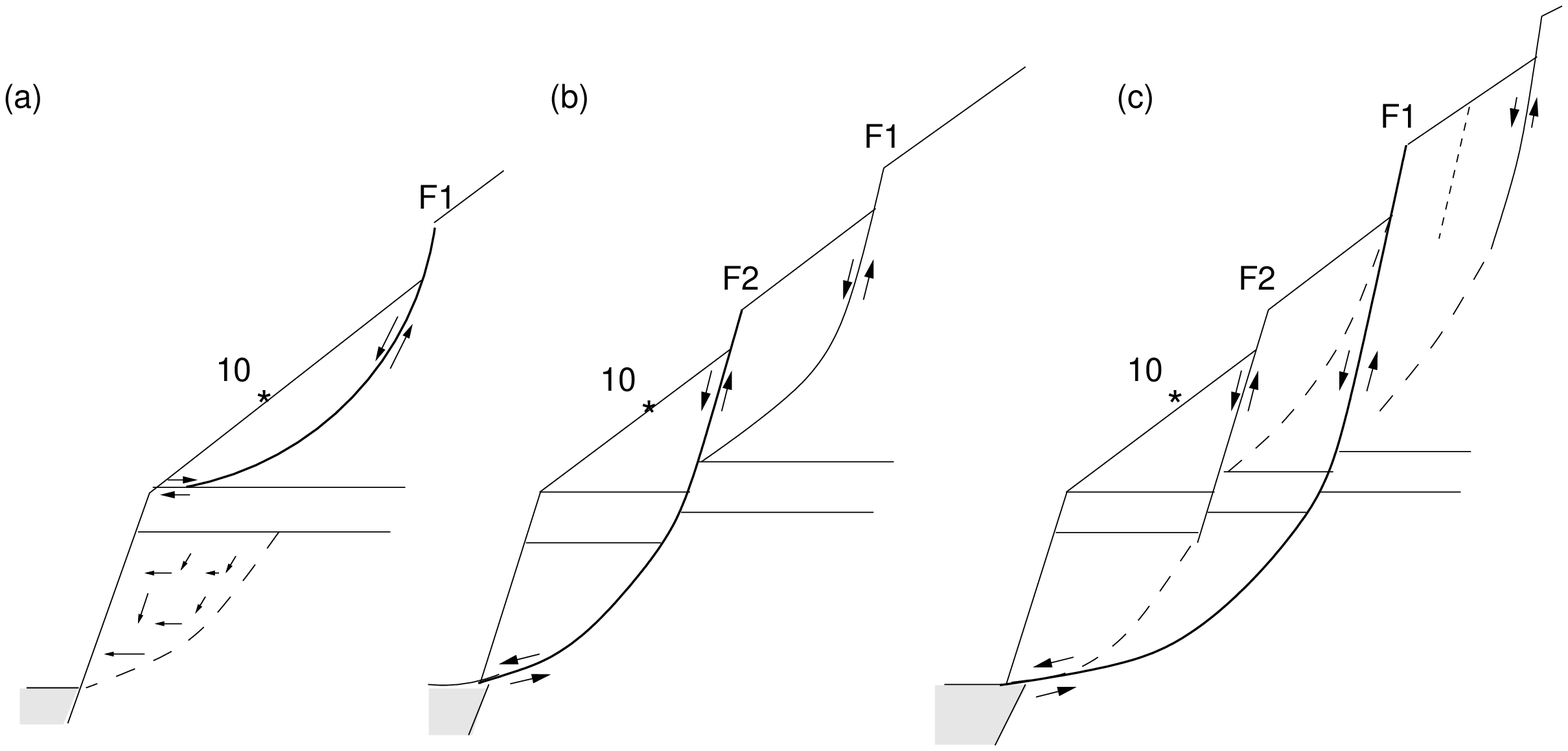,width=16cm}
\caption{
Schematic structural interpretation of one possible mechanism involved in
the 1986-1988 crisis. The 3 schematic cross sections are the proposed landslide
geometries, before 1986 (a), during the 1987 acceleration (b), and 
after 1988 (c).
{\it Follacci et al.} [1993] argue for  the failure of the strong 
gneiss bed (F2 fault)
in the NW block as the driving force behind the 1986-1987 
accelerating phase (b).
In the same period, the development of the upper NW crack, (F1 fault on central
cross section), that released the landslide from its head driving force,
appears as the key parameter to slow down the accelerating slide.
{\it Guglielmi and Vengeon.} [2002] argue for all the surface faulting patterns
to converge  at shallow depth   as listric faults that define a decollement
level which is the sliding surface.
The star shows the location of benchmark 10 (adapted from [{\it 
Follacci et al.}, 1993]).}
\label{coupefollacciorguglielmi}
\end{figure}

In 1985-1986, a transverse crack initiated in the upper part of the
NW block. It reaches 50 m of vertical offset in 1989. The maximum
rate of change of the
fracture size and of its opening occurred in 1987 [{\it Follacci et
al.}, 1993].
This new transverse crack uncoupled the NW block from the upper part of the
mountain, which moved at a much smaller velocity below 1 mm/day since 1985-86
[{\it Follacci et al.}, 1993] (Figures  \ref{photo-clapieres} and
\ref{coupefollacciorguglielmi}).
Since summer 1988, an homogenization of the surface morphological faces and a
regression of the main summit scarp were reported. The regression of the summit
scarp was observed as a new crack started to open in September 1988. Its length
increased steadily to reach 500 m and its width reached 1.75 m in
November 1988.
Accordingly, the new elevation of main scarp in the SE block reaches 1780 m.
This crack, which defined a new entity, that is the
upper SE block, has remained locked since then (Figures
\ref{photo-clapieres} and
\ref{coupefollacciorguglielmi}).

\subsubsection {Current understanding of La Clapi\`ere acceleration}

On the basis of these observations and simple numerical models, an
interpretative model for the 1986-1988 regime change was proposed by
{\it Follacci et al.}, [1993] [see also for a review {\it Susella and
Zanolini}, 1996].
In fact, these models do not explain the origin of the acceleration but rather
try to rationalize kinematically the different changes of velocity  and
why the acceleration did not lead to a catastrophic sliding but re-stabilized.
The reasoning is based on the fact that the existing and rather
strong correlation
between the river flow in the valley at the bottom
and the slope motion  (see Figure \ref{debitpluie})
is not sufficient to explain both the de-stabilizing phase
and its re-stabilization. This strongly suggests that the
hydrological conditions
are not the sole control parameters explaining both the strong 1986-1987
accelerating and the equally strong slowdown in 1988-1990.

{\it Follacci et al.} [1988, 1993] argue that the failure of the strong gneiss
bed in the NW block was the main driving force of the acceleration in
1986-1987.
According to this view, the failure of this bed induced changes in both the
mechanical boundary conditions and in the local hydro-geological setting
(Figure \ref{coupefollacciorguglielmi}). Simultaneously, the development of the
upper NW crack, that freed the landslide from its main driving force, appears
as a key parameter to slow down the accelerating slide. The hypothesized
changes in hydrological boundary conditions can further stabilize the
slide after the 1986-1987 transient acceleration.

Several works have attempted to fit the velocity time series of La
Clapi\`ere landslide and predict its future evolution, using a framework
similar to the Vaiont landslide discussed above.
The displacement of different benchmarks over the 1982-1986
period has been analysed. An exponential law has been fitted to the
1985-1986 period
[{\it Vibert et al.}, 1988].
Using the exponential fit and a failure criterion that the landslide
will collapse when
the velocity reaches a given threshold, the predicted collapse time
for the landslide
ranges from 1988 for NW benchmark to 1990 for the SE benchmarks.
Plotting the inverse of the velocity as a function of time as in
(\ref{jmgjkla}) has
been tried, hoping  that this law holds with $\alpha=2$ providing a
straightforward
estimation of $t_c$. This approach applied to La Clapi\`ere velocity
data predicts a
collapse in 1990 for the upper NW part and in 1988-1989 for the SE
part of the landslide.
To remove the fluctuations of the velocity induced by changes in river flow,
an ad-hoc weighting of the velocity data was used by [{\it Vibert et
al.}, 1988].
An attempt to more quantitatively estimate the relation
between the river flow and the landslide velocity was proposed by
{\it Rat} [1988].
{\it Rat} [1988] stresses the importance of removing the fluctuations
of the velocity
induced by changes in the river flow before any attempt to predict
the collapse time.

\subsection{Analysis of the cumulative displacement and velocity data
with the slider-block model}

\subsubsection{ La Clapi\`ere sliding regime: 1982-1987}

We fit the monthly measurements of the displacement of several representative
benchmarks  with the slider-block friction model.
In the sequel, we will show results for benchmark 10 which is
located in the central part of the landslide (Figure \ref{photo-clapieres}),
and which is representative of the average landslide behavior during the
1982-1995 period [{\it Follacci}, personal communication 2001].
We have also obtained similar results for benchmark 22.

\begin{figure}
\psfig{file=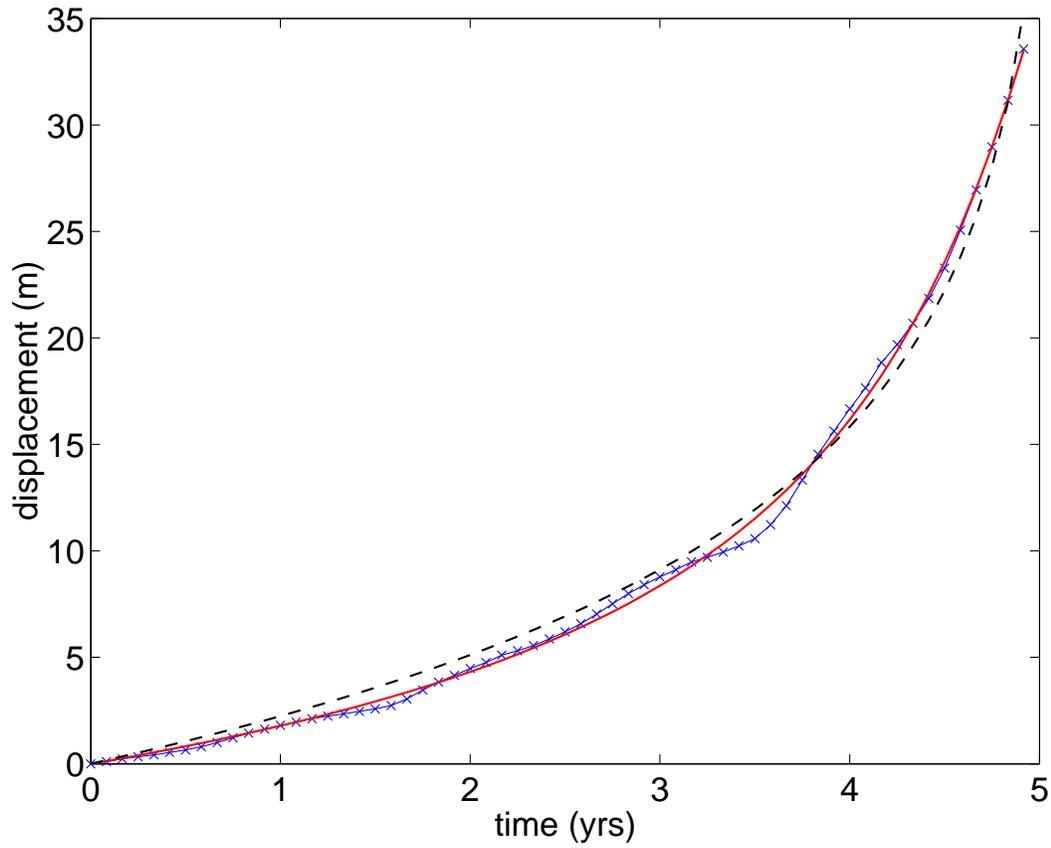,width=14cm}
\caption{ Displacement for benchmark 10 of la Clapi\`ere landslide (crosses)
and fit using the friction model.
The best fit gives $m=0.98$ (black line). The dashed line shows the best fit
obtained when imposing $m=1.5$ for comparison.}
\label{depclap}
\end{figure}

We consider only the accelerating phase in the time interval
$[1982.9; 1987.9]$.
As for the Vaiont landslide, the inversion provides the values of the
parameters $m$, $T$, $D$, and the initial condition $x_i$ of the
state variable.
For La Clapi\`ere, we analyze the displacement as
it has a lower noise level compared with the
velocity. In the Vaiont case, the data is of sufficiently good quality to
use the velocity time series which allows us to compare with previous studies.
The best fit to the displacement of benchmark 10 is shown in Figure
  \ref{depclap}. The model parameters are $m=B/A$=0.98 and
the  initial value of the reduced state variable is $x_i=39$.
While $m$ is very close to one, the value of $x_i$ significantly larger
than 1 argues for La Clapi\`ere landslide to be in the stable regime
(see Figure \ref{classregime} and Table 1).
Similar results are obtained for the other benchmarks.
Since the landslide underwent different regimes, it is important to perform
these inversions for different time periods, that is, the fits are done
from the first measurement denoted time $t=0$ (year $1982.9$)
to a later $t=t_{\rm max}$, where $t_{\rm max}$ is increased from approximately
$2$ years to $5$ years after the initial starting date.
This last time $t \approx 5$ years (end of 1987)
corresponds to the time at which the slope velocity reached its peak.
For all inversions except the first two  point with $t_{\rm max}\approx 2$ yrs,
the best fit always select an exponent  $0<m<1$ and an initial state variable
$x_i \gg 1$, corresponding to a stable asymptotic sliding without finite-time
singularity. For $t_{\rm max}<4$ years (that is, using data before the end
  of 1986), a few secondary best solutions are found with very different values,
from $m=-3000$ to $m=29$, indicating that $m$ is poorly constrained.
We have also performed sensitivity tests using synthetic data sets
generated with
the friction model with the same parameters as those obtained for La
Clapi\`ere.
These tests show that a precise determination of $m$ is impossible
but that the inversion recovers the true regime $m<1$.

The transition time (defined by the inflection point of the velocity)
is found to
increase with $t_{max}$.
This may argue for a change of regime from an acceleration regime to a
restabilization before the time $t=1988$ of the velocity peak.
The parameters $S$ and $x_i$ are also poorly
constrained. Similar results are obtained for different benchmarks as well as
when fitting the velocity data instead of the
displacement [{\it Helmstetter}, 2002; {\it Sornette et al.},2003].
 The velocity data show large fluctuations,
in part due to yearly fluctuations of the precipitations.
The inversion is therefore even  more unstable than the inversion of the
displacement, but almost all points give $m<1$ and $x_i>1$.
Such fluctuations of the inverted solution may indicate that the use
of constant
friction parameters  to describe a period where 2 regimes interact,
i.e., an accelerating  phase up to 1987 followed by a decrease in
sliding rate since 1988,  does not describe adequately the landslide behavior
for the whole time period 1982-1996. Observed changes in morphology
as suggested in Figures \ref{photo-clapieres} and
\ref{coupefollacciorguglielmi} provide evidence for changes both in
driving forces
and in the geometry of the landslide, including possible new sliding surfaces.

\subsubsection{ La Clapi\`ere decelerating phase: 1988-1996}
The simple rigid block model defined with a single block
and with velocity and state dependent friction law
cannot account for what happened after the velocity peak, without invoking
additional ingredients. Departure from the model prediction can be used
as a guide to infer in-situ landslide behavior.
Recall that, during the interval 1988-1990, the monthly recorded
velocities slowed down to velocity 6 times smaller than the 1987 peak values.
This deceleration cannot be explained with the friction model using constant
friction parameters. Indeed, for $B/A=m<1$, under a constant geometry and fixed
boundary conditions, the velocity increases and then saturates at its 
maximum value.
In order to explain the deceleration of the landslide, a
change of material properties can be invoked
(embodied for example in the parameter $m=B/A$)
or a change of the state variable $\theta$ that describes the duration of
frictional contacts, maybe due to a change in the sliding surfaces.

We have not attempted in this study to fit both the accelerating and
the decelerating phases with the slider-block model due to the large number
of free parameters it will imply relatively to the small number
of points available. Further modeling would allow block partitioning,
fluctuations of the slope angle and change with time of the friction 
parameters.
Our purpose is here to point out how different landsliding regimes can be
highlighted by the introduction of a velocity and state friction law in
this basic rigid block model.

\section{Discussion and conclusion}

We have presented a quantitative analysis of the displacement 
history for  two landslides, Vaiont and La Clapi\`ere, using
a slider-block friction model.
An innovative concept proposed here was to apply to landslides
the state and velocity dependent friction law established in the laboratory
and used to model earthquake friction. Our inversion of this simple
slider-block friction model shows that the observed movements can be well
reproduced with this simple model and suggest the Vaiont landslide
(respectively La Clapi\`ere landslide) as belonging to the velocity
weakening unstable (respectively strengthening stable) regime.
Our friction model assumes that the material properties
embodied in the key parameters $m=B/A$ and/or the initial value
of the state variable of the friction law control the sliding regime.

Our purpose was here to point out how different landsliding regimes can be
highlighted by the introduction of a velocity and state friction law in
a basic rigid block model.
Even if the displacement is not homogeneous for the two landslides,
the rigid block model provides a good fit to the observations and a first
step towards a better understanding of the different sliding regimes and
the potential for their prediction.

For the cases studied here, we show that a power law increase with time of
of the slip velocity can be reproduced by a rigid slider block model.
This first order model rationalizes the previous empirical law
suggested by  Voight [1988]. Following  {\it Petley et al.} [2002], we
suggest that the landslide power law acceleration 
emerges in the presence of a rigid block, i.e., this 
corresponds to the slide of a relatively stiff material.
 {\it Petley et al.} [2002]  report that, for some other types of landslides
in ductile material, the slips do not follow a linear dependence with 
time of the inverse landslide velocities.  They suggest that the
latter cases are reminiscent of the signature of landsliding associated with
a ductile failure in which crack growth does not occur.
In contrast, they proposed that the linear dependence of the
inverse velocity of the landslide  as a function of time
is reminiscent of crack propagation, i.e., brittle deformation
on the basal shear plane.  Our contribution suggests that friction
is another possible process that
can reproduce the same accelerating pattern than the one proposed
to be driven by crack growth on a basal shear plane [{\it Petley et al.},
2002; {\it Kilburn and Petley}, 2003].   The friction model used in our 
study requires the existence of an
interface. Whether this friction law should change for ductile material
is not clear. The lack of direct observations of the shearing zone and its
evolution through time makes difficult
the task of choosing between the two classes of models,
crack growth versus state-and-velocity-dependent friction. 
We recover here the still on-going debate for earthquakes,
which can be seen as either frictional or faulting events. 

For the Vaiont landslide, this physically-based model suggests that 
this landslide was in the unstable regime. 
For La Clapi\`ere landslide, the inversion of the displacement data
for the accelerating phase 1982-1887 up to the maximum of the velocity
gives $m<1$, corresponding to the stable regime.
The deceleration observed after 1988 implies that, not only
is La Clapi\`ere landslide in the stable regime but in addition,
some parameters of
the friction law have changed, resulting in a change of sliding regime
from a stable regime to another one characterized by a smaller velocity,
as if some stabilizing process or reduction in stress was occurring.
Possible candidates for a change in landsliding regime include the average
dip slope angle, the partitioning of blocks, new sliding surfaces and changes
in interface properties.
The major innovation of the frictional slider-block model
which is explored further in [{\it Sornette et al.}, 2003] is to embody the
two regimes (stable versus unstable) in the same physically-based 
framework, and to offer a way of distinguishing empirically between 
the two regimes, as shown by our analysis of the two cases provided 
by the Vaiont and La Clapi\`ere landslides.

\acknowledgments
We thanks C. Scavia and Y. Guglielmi for key supports to capture 
archive data for
  Vaiont and La Clapi\`ere  landslide respectively.
  We are very grateful to N. Beeler, J. Dieterich, Y. Guglielmi,  D. Keefer,
J.P. Follacci, J.M. Vengeon  for useful suggestions and discussions.
AH and JRG were supported by INSU french grants, Gravitationnal 
Instability ACI.
SG and DS acknowledge support from the James S. Mc Donnell Foundation 21st
century scientist award/studying complex systems.

\appendix
\section{Appendix A: Derivation of the full solution of the frictional problem}

We now provide the full solution of the frictional problem.
First, we rewrite (\ref{vxcxxzx}) as
\be
\dot{\delta} = S~ D_c ~
\left({\theta \over \theta_0}\right)^{-m}~,
\label{ngfvw}
\ee
where
\be
S \equiv {\dot{\delta}_0 ~ e^{{\tau \over \sigma}-\mu_0 \over A} \over D_c}
\label{mmjdl}
\ee
and
\be
m \equiv {B \over A}~.
\label{mgmdl}
\ee
Putting (\ref{ngfvw}) in (\ref{qwertyu}) gives
\be
{{\rm d}(\theta /\theta_0) \over {\rm d}t} = {1 \over \theta_0}
- S~ (\theta /\theta_0)^{1-m}~.
\label{qwerssssstyu}
\ee
The case $m=1$ requires a special treatment since the dependence in
$\theta$ disappears in the right-hand-side of (\ref{qwerssssstyu})
and ${{\rm d}\theta \over {\rm d}t}$ is constant.

For $m \neq 1$, it is convenient to introduce the reduced variables
\be
x \equiv (S \theta_0)^{1/(1-m)}~{\theta \over \theta_0}~,
\label{mgmbm}
\ee
and
\be
D \equiv D_c~\left(S \theta_0^m \right)^{1 \over 1-m}~.
\label{hfff}
\ee
Then, (\ref{ngfvw}) reads
\be
{\dot{\delta} \over \dot{\delta}_0} = D ~x^{-m}~.
\label{velocity}
\ee
Putting (\ref{ngfvw}) in (\ref{qwertyu}) to eliminate the dependence
in $\dot{\delta}$, we obtain
\be
{{\rm d}x \over {\rm d}t'} = 1 - x^{1-m}~,
\label{ode}
\ee
where $t' = t/T$ with
\be
T = {D_c \over D} =
\left[{D_c \over \dot{\delta}_0 \theta_0^m}\right]^{1/(1-m)}~
e^{{\tau \over \sigma}-\mu_0 \over B-A}~.
\label{mgmlla}
\ee
In the sequel, we shall drop the prime and use the dimensionless
time $t'$, meaning that time is expressed in units of $T$ except
stated otherwise.

The block sliding behavior is determined by first solving the
equation (\ref{ode}) for the normalized state variable $x(t)$ and then
by inserting this solution in (\ref{velocity}) to get the slip velocity.
Equation (\ref{ode})
displays different regimes as a function of $m$
and of the initial value $x_i$ compared to $1$ that we now classify.

\subsection{Case $m=B/A>1$}

For $m>1$ and $x_i<1$,
the initial rate of change ${{\rm d}x \over {\rm d}t}$
of the state variable is negative. The initial decay of $x$ accelerates
with time and $x$ reaches $0$ in finite time. Expression (\ref{velocity}) shows
that  $\delta(t)$ continuously accelerates and reaches infinity in finite time.
Close to the singularity, we can neglect the first term $1$ in the
right-hand-side of
(\ref{ode}) and we recover the asymptotic solution
(\ref{nglnwq},\ref{tc},\ref{ngaaz}):
\be
x(t) \simeq  m^{1 \over m} ~(t_c-t)^{1 \over m} ~,
\label{mmsls}
\ee
where the critical time $t_c$ is determined by the initial condition
$x(t=0)=x_i$
\be
t_c={x_i^m \over m}~.
\label{tcbis}
\ee

For $m>1$ and $x_i>1$,  the initial rate of change
${{\rm d}x \over {\rm d}t}$ of the state variable is positive, thus
$x$ initially
increases. This growth goes on, fed by the positive feedback embodied
in (\ref{ode}).
At large times, $x$ increases asymptotically at the constant rate
${{\rm d}x \over {\rm d}t}=1$ leading to $x(t) \approx t$.
Integrating equation (\ref{velocity}) gives
\be
\delta(t) = \delta_{\infty} - {\dot{\delta}_0 D \over m-1}~{1 \over t^{m-1}}~,
\label{mgmmlls}
\ee
at large times.
The asymptotic value of the displacement $\delta_{\infty}$ is determined
by the initial condition. This regime thus describes a decelerating slip
slowing down as an inverse power of time. It does not correspond to a
de-stabilizing landslide but to a power law plasticity hardening.

\subsection{Case $m=B/A=1$}

In this case, the variables (\ref{mgmbm}) and (\ref{hfff}) are
not defined and we go back to (\ref{qwertyu}) (which uses the unnormalized
state variable $\theta$ and time $t$) to obtain
\be
{{\rm d} \theta \over {\rm d} t} = 1-S \theta_0~,
\ee
where $S$ is defined by (\ref{mmjdl}) and depends on the material properties
but not on the initial conditions. If $S \theta_0>1$, $\theta$ decays
linearly and reaches $0$ in finite time. This retrieves the finite-time
singularity, with the slip velocity diverging as $1/(t_c-t)$ corresponding
to a logarithmic singularity of the cumulative slip.
If $S \theta_0<1$, $\theta$ increases linearly with time. As a consequence,
the slip velocity decays as $\dot{\delta} \sim 1/t$ at large times
and the cumulative slip grows asymptotically logarithmically
as $\ln t$. This corresponds to a standard plastic hardening behavior.

\subsection{Case $m=B/A<1$}

For $x_i>1$, the initial rate of change ${{\rm d}x \over {\rm d}t}$
of the state variable is negative, thus $x$ decreases and converges to the
stable fixed point $x=1$ exponentially as
\be
x=1+ a e^{-{t \over t^*}} ~,
\label{mi11}
\ee
   where the relaxation time $t^*$ is given by
\be
t^*={1 \over 1-m}
\label{tau}
\ee
in units of $T$ and $a$ is a constant determined by the initial
condition. Starting
from some initial value, the slip velocity increases for $0<m<1$
(respectively decreases for $m<0$) and
converges to a constant, according to (\ref{ngfvw},\ref{velocity}).

For $x_i<1$, the initial rate of change ${{\rm d}x \over {\rm d}t}$
of the state variable is positive, and $x$ converges exponentially
toward the asymptotic stable fixed point $x=1$. As $\theta$
increases toward a fixed value, this implies that the slip velocity
decreases for $0<m<1$
(respectively increases for $m<0$) toward a constant value.

\end{article}

\clearpage



\clearpage

\clearpage

\clearpage

\clearpage

\begin{figure}
\caption{a) See figure in jpeg format.
Picture of La Clapi\`ere landslide taken in 1979.
The volume of mostly gneiss rocks implied in the landslide
is estimated to be around $50\times 10^6$ m$^3$.
been monitored
The summit scarp are not connected.
b) Picture of La Clapi\`ere landslide taken in 1999.
The global surfacial pattern is preserved.
The main feature related to the 1982-1988 crisis is a new summit scarp with
a total displacement of about 100 m in 1999, indicated by an arrow in 
figure (b).
}
\label{photo-clapieres}
\end{figure}

\clearpage

\begin{figure}
\psfig{file=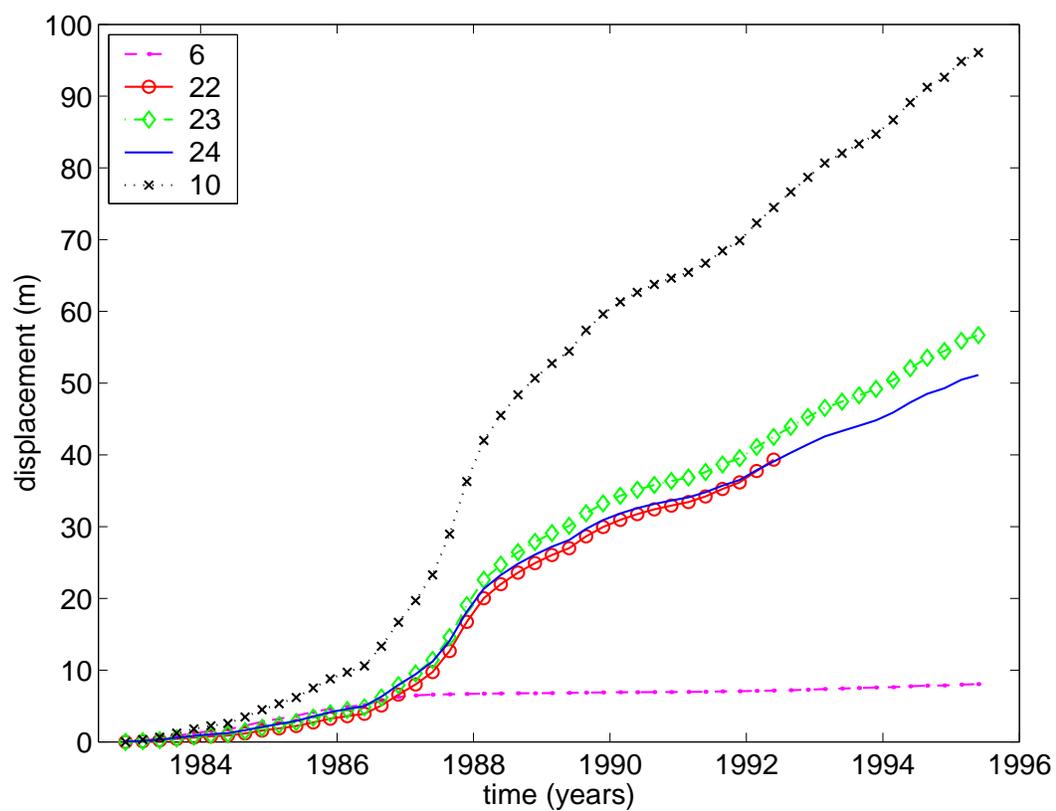,width=14cm}
\caption{Displacement for the 5 benchmarks on La Clapi\`ere  site shown
in Figure  \ref{photo-clapieres}.}
\label{4benmarkclap}
\end{figure}

\clearpage

\clearpage

\clearpage

\begin{table}
\caption{
Synthesis of the different regimes of slip as a function
of $m=B/A$ (by definition (\ref{mgmdl})), of the initial condition
$x_i$ on $\theta$ and of the material parameter $S$ defined by 
(\ref{mmjdl}). $A$ and $B$ are defined in (\ref{vxcxxzx})
and are determined by material properties. $x_i$ is the initial value of the
reduced state variable $\theta$ defined in (\ref{mgmbm}). FTS stands
for ``finite-time singularity.'' The parenthesis ($x_i$) and ($S$) in 
the first column indicates which is the control parameter determining 
the nature of the slip. The parameter $S$ is independent of the initial 
conditions. While $A$ is always found positive in laboratory experiments, 
negative $B$-values are sometimes found [{\it Blanpied et al.}, 1995] 
leading to the possibility of having $m<0$: this rather special case
corresponds to a friction coefficient decreasing with the
increase of the surface of contacts.
}
\label{table}
\begin{center}
\renewcommand{\arraystretch}{1.4}
\setlength\tabcolsep{6pt}
\begin{tabular}{@{}llll}
\hline\noalign{\smallskip}
   & $x_i, S<1$ & $x_i, S>1$  \\

\hline
$m>1$ ($x_i$) & FTS (\ref{nglnwq},\ref{tc},\ref{ngaaz}) &
power law plasticity hardening (\ref{mgmmlls})\\
$m=1$ ($S$) & $\dot{\delta} \sim 1/t$ and $\delta \sim \ln t$ &
FTS (\ref{nglnwq},\ref{tc},\ref{ngaaz}) \\
$0 < m < 1$ ($x_i$) & $\theta \downarrow$ const, $\dot{\delta}
\uparrow$ const &
$\theta \uparrow$ const, $\dot{\delta} \downarrow$ const \\
$m<0$ ($x_i$)& $\theta \downarrow$ const, $\dot{\delta} \downarrow$ const &
$\theta \uparrow$ const, $\dot{\delta} \uparrow$ const \\
\noalign{\smallskip}
\hline
\noalign{\smallskip}
\end{tabular}
\end{center}
\end{table}
\end{document}

%% file: update.tex
\makeatletter
\let\reset@font\empty
\makeatother

%% file: Laclapiererevised12.bbl
\begin{references}

\reference Anifrani, J.-C., C. Le Floch, D. Sornette and B. Souillard,
Universal Log-periodic correction to renormalization group scaling
for rupture stress
prediction from acoustic emissions, {\it J. Phys. I France 5}, 631-638, 1995.

\reference Bhandari, R. K, Some lessons in the investigation and 
field monitoring
of landslides, {\it Proceedings  5th Int. Symp. Landslides Lausanne 1988, eds
C.  Bonnard , 3}, 1435-1457, Balkema, 1988.

\reference Blanpied, M. L., D. A. Lockner and J. D. Byerlee,
Frictional slip of granite at hydrothermal conditions,
{\it J. Geophys. Res., 100}, 13045-13064, 1995.

\reference Broili, L., New knowledge on the geomorphology of the 
Vaiont slide slip
surfaces, {\it Rock mechanics and Engineering, 5}, 38-88, 1967.

  \reference Campbell, C.S., Self-lubrification for long
runout landslides, {\it Journal of Geology, 97}, 653-665, 1989.

\reference Campbell, C.S.,
   Rapid granular flow,  {\it Annu. Rev. of Fluid Mech., 22}, 57-92, 1990.

\reference Caplan-Auerbach, J., C. G. Fox and F. K. Duennebier,
Hydroacoustic detection of submarine landslides on Kilauea volcano,
  {\it Geophys. Res. Letts, 28}, 1811-1813, 2001.


\reference David, E. and ATM, Glissement de La Clapi\`ere, St Etienne 
de Tin\'ee,
Etude cin\'ematique, g\'eomorphologique et de stabilit\'e, {\it rapport CETE
Nice}, France, 88 pp., 2000.

\reference Davis, R.O., N. R. Smith and G. Salt, Pore fluid 
frictional heating and
stability of creeping landslides, {\it  Int. J. Num. Anal. Meth.
Geomechanics, 14},  427-443, 1990.


\reference Dieterich, J., Time dependent friction and the mechanics 
of stick slip,
   {\it Pure Appl. Geophys., 116}, 790-806, 1978.

\reference Dieterich, J. H., Earthquake nucleation on faults with
rate- and state-dependent strength, {\it Tectonophysics, 211}, 115-134, 1992.

\reference Durville, J.L., Study of mechanisms and modeling of large 
slope movements,
{\it  Bull. Int. Ass. Engineering Geology, 45}, 25-42, 1992.

\reference Eisbacher, G.H., Cliff collapse and rock avalanches in the Mackenzie
Mountains, Northwestern Canada, {\it Can. Geot. J., 16}, 309-334, 1979.

\reference Erismann, T.H. and G. Abele, {\it Dynamics of Rockslides 
and Rockfalls},
Springer, 300 pp, 2000.

\reference Follacci, J. P.,  Photographic album of La Clapi\`ere landslide,
  {\it CETE M\'editerran\'ee}, 2000.

\reference Follacci, J. P., P. Guardia and J. P. Ivaldi,
La Clapi\`ere landslide in its geodynamical setting,
{\it Bonnard eds , Proc. 5th Int. Symp. on Landslides, 3}, 1323-1327, 1988.

\reference Follacci, J.-P., L. Rochet and  J.-F. Serratrice, 
Glissement de La Clapi\`ere,
St Etienne de Tin\'ee, Synth\`ese des connaissances et actualisation des
risques, {\it rapport 92/PP/UN/I/DRM/03/AI/01, Minist\`ere Environnement,  76}
pp., 1993.

\reference Fruneau, B., J. Achache and C. Delacourt, Observation and
modeling of the Saint-Etienne-de-Tin\'ee landslide using SAR interferometry,
{\it Tectonophysics, 265}, 181-190, 1996.

\reference Gluzman, S. and D. Sornette,
Self-Consistent theory of rupture by progressive diffuse damage,
{\it Physical Review E, 6}, 306 N6 PT2:6129,U241-U250, 2001.

\reference Gomberg, J., P. Bodin, W. Savage  and M. E. Jackson,
Landslide faults and tectonic faults, Analogs? -- The slumgullion
earthflow, Colorado, {\it Geology, 23}, 41-44, 1995.

\reference Gomberg, J., N. Beeler and M. Blanpied,
On rate-state and Coulomb failure models, {\it J. Geophys. Res., 105},
7857-7871, 2000.

\reference Guglielmi, J., and J. M. Vengeon,
Interrelation between gravitational patterns and structural fractures
La Clapi\`ere, French Alps, submitted to {\it Geomorphology}, 2002.

\reference Helmstetter, H.,
Rupture and Instabilities: seismicity and landslides,
{\it PhD thesis}, Grenoble University, pp 387, 2002.

\reference Heim A., {\it Bergsturz and Menschenleben,  Zurich}, 1932.

\reference Hendron, A. J.  and F. D. Patton,
The Vaiont slide, a geotechnical analysis based on new geologic observations
of the failure surface, {\it US Army Corps of Engineers Technical
Report GL-85-5 (2 volumes)}, 1985.

\reference Hoek, E. and E. T. Brown, {\it Underground excavation in rock},
Institution of Mining and Metallurgy, London, 1980.

\reference Hoek, E., and J. W. Bray, {\it Rock slope engineering}, 
3rd Edn (rev),
Institution of Mining and Metallurgy and E\&FN Spon, London, pp 358, 1997.

\reference Jaume, S. C. and L. R. Sykes,
Evolving toward a critical point: A review of accelerating seismic
moment/energy release prior to large and great earthquakes, {\it Pure
and Appl. Geophys., 155}, 279-305, 1999.

\reference Kennedy B. A. and K. E.  Niermeyer, Slope Monitoring 
systems used in the
Prediction of a Major Slope Failure at the Chuquicamata Mine,
Chile, {\it Proc. on  Planning Open Pit Mines, Johannesburg}, Balkema,
215-225, 1971.

\reference Kilburn, C. R. J., and D. N. Petley,
Forecasting giant, catastrophic slop collapse: lessons from Vajont,
  Northern Italy,  in press. {\it Geomorphology}, 2003.

\reference Korner, H. J., Reichweite and Geschwindigkeit von Bergsturzen und
Fleisscheneelawinen, {\it Rock mechanics, 8}, 2256-256, 1976.

\reference Malet, J. P., O. Maquaire and E. Calais,
The use of Global Positioning System techniques for the continuous
monitoring of landslides: application to the
Super-Sauze earthflow (Alpes-de-Haute-Provence, France),
{\it Geomorphology, 43}, 33-54, 2002.

\reference Mantovani, F., R. Soeters and C. J. Vanwesten,
Remote sensing techniques for landslide studies and hazard zonation in Europe,
{\it Geomorphology, 15}, 213-225, 1996.

\reference Marone, C., Laboratory-derived friction laws and their 
application to seismic
faulting, {\it Ann. Rev. Earth Planet. Sci., 26}, 643-696, 1998.

\reference Muller L., The rock slide in the Vaiont valley, {\it 
Felsmechanik und
Ingenoirgeologie, 2 (3-4)}, 148-212, 1964.

\reference Muller L., News consideration on the Vaiont slide, 
{\it Felsmechanik und Ingenoirgeologie, 6}, 1-91, 1968.

\reference Parise, M.,
Landslide mapping techniques and their use in the assessment of
the landslide hazard, {\it Phys. Chem. Earth, Part C- Solar-Terr. Plan.
Sci.,  26}, 697-703, 2001.

\reference Petley, D.N., Bulmer, M.H. and Murphy W. ,
Patterns of movement in rotational and translational landslides, 
Geology. 30(8), 719-722, 2002.

\reference Rabotnov, Y. N., {\it Creep problems in Structural Members},
{\it North-Holland eds}, Amsterdam, 1969.

  \reference Rat, M., Difficulties in forseeing failure in landslides 
- La Clapi\`ere, French Alps,
{\it Proceedings  5th Int. Symp. Landslides Lausanne 1988, eds C. Bonnard,
vol 3}, 1503-1504, Balkema, 1988.

\reference Rousseau, N.,
Study of seismic signals associated with rockfalls at 2 sites on the
Reunion island (Mahavel Cascade and Souffri\`ere cavity),
{\it PhD thesis}, IPG Paris, 1999.

\reference Ruina, A. L., Slip instability and sate variable friction laws,
	 {\it J. Geophys. Res., 88}, 10359-10370, 1983.

\reference Saito, M., Forecasting the Time of occurrence of a Slope Failure,
{\it Proc. 6th Int.  Conf. Soil Mech. \& Found. Eng., Montreal, 
vol.2}, 537-541, 1965.

\reference Saito, M., Forecasting time of Slope Failure by Tertiary Creep,
{\it Proc. of 7th  Int. Conf. Soil Mech. \& Found. Eng.}, Mexico City, vol. 2,
677-683, 1969.

\reference Saito, M. and H. Uezawa, Failure of soil due to creep, 
{\it Proc of 6th
Int. Conf.  Soil Mech. \& Found. Eng., Montreal, vol. 1}, 315-318,  1961.

\reference Sammis, S. G. and D. Sornette,
Positive Feedback, Memory and the Predictability of Earthquakes,
{\it Proceedings of the National Academy of Sciences USA, 99}, 2501-2508, 2002.

\reference Scholz, C. H., The mechanics of earthquakes and faulting
{\it Cambridge University Press}, 1990.

\reference Scholz, C. H., Earthquakes and friction laws, {\it Nature, 
391}, 37-42, 1998.
 

\reference Sornette, D.,
Predictability of catastrophic events: material rupture, earthquakes,
turbulence, financial crashes and human birth,
{\it Proceedings of the National Academy of Sciences USA, 99}, 2522-2529, 2002.

\reference Sornette, D., A. Helmstetter, J. V. Andersen, S. Gluzman, 
J.-R. Grasso,
V. Pisarenko, Towards landslide predictions: two case studies, 
submitted to Phys. Rev. E, 2003.

\reference Sornette, D. and C. G. Sammis, Complex critical
exponents from renormalization group theory of earthquakes:
Implications for earthquake predictions, {\it J. Phys. I France, 5}, 
607-619, 1995.

\reference Susella, G. and F. Zanolini,  Risques g\'en\'er\'es par les grands
mouvements de terrains,  {\it eds, Programme Interreg 1, France-Italie},
207 pp., 1996.

\reference Van Asch, T. W. J., J. Buma and L. P. H. Van Beek,
A view on some hydrological triggering systems in landslides,
{\it Geomorphology, 30}, 25-32, 1999.

\reference Vangenuchten, P. M. B. and H. Derijke,
Pore water pressure variations causing slide velocities and
accelerations observed in a seasonally active landslide,
{\it Earth Surface Processes and Landforms, 14}, 577-586, 1989.


\reference Vibert C., M. Arnould, R. Cojean, J. M. Cleac'h,
   An attempt to predict the failure of a mountainous slope at St 
Etienne de Tin\'ee, France,
{\it Proceedings  5th Int. Symp. Landslides Lausanne 1988, eds C. Bonnard,
vol 2}, 789-792, Balkema, 1988.

\reference  Voight, B. {\it eds, 2, Engg.
Sites, Development  in Geotech. Engg., vol. 14b}, 595-632, 1978.

\reference Voight, B., A method for prediction of Volcanic Eruption,
{\it Nature, 332}, 125-130, 1988.

\reference Voight, B. A., A relation to describe rate-dependent 
material failure,
{\it Science, 243}, 200-203, 1989.

\reference Voight B., Materials science laws applies to time forecast of slope
failure, {\it Proceedings  5th Int. Symp. Landslides Lausanne 1988, eds
C. Bonnard, vol 3},  1471-1472, Balkema, 1988.


\reference Xu Z. Y., S. Y. Schwartz and T. Lay, Seismic
wave-field observations at a dense small-aperture array
located on a landslide in the Santa Cruz Mountains, California,
{\it Bull. Seis. Soc. Am., 86}, 655-669, 1996.

\end{references}
